\documentclass[11pt]{article}
\usepackage{hyperref}
\pdfoutput=1
\begin{document}
\title{Slow dynamics of phospholipid monolayers \\ at the air/water interface}
\author{Siyoung Q. Choi and Todd M. Squires \\
\\\vspace{6pt} Department of Chemical Engineering
, \\ University of California, Santa Barbara, CA 93106, USA}
\maketitle
\begin{abstract}
Phospholipid monolayers at the air-water interface serve as model systems for various biological interfaces, e.g. lung surfactant layers and outer leaflets of cell membranes. Although the dynamical (viscoelastic) properties of these interfaces may play a key role in stability, dynamics and function, the relatively weak rheological properties of most such monolayers have rendered their study difficult or impossible. A novel technique to measure the dynamical properties of fluid-fluid interfaces have developed accordingly. We microfabricate micron-scale ferromagnetic disks, place them on fluid-fluid interfaces, and use external electromagnets to exert torques upon them. By measuring the rotation that results from a known external torque, we compute the rotational drag, from which we deduce the rheological properties of the interface. Notably, our apparatus enable direct interfacial visualization while the probes are torqued. 

In this fluid dynamics video, we directly visualize dipalmitoylphosphatidylcholine(DPPC) monolayers at the air-water interface while shearing. At about 9 mN/m, DPPC exhibits a liquid condensed(LC) phase where liquid crystalline domains are compressed each other, and separated by grain boundaries. Under weak oscillatory torque, the grain boundaries slip past each other while larger shear strain forms a yield surface by deforming and fracturing the domains. Shear banding, which is a clear evidence of yield stress, is visualized during steady rotation. Remarkably slow relaxation time was also found due to slow unwinding of the stretched domains. 

\end{abstract}

Two videos are (\href{http://ecommons.library.cornell.edu/bitstream/1813/14115/3/gallery_of_fluid_motion_final_low.mpg}{low quality}, \href{http://ecommons.library.cornell.edu/bitstream/1813/14115/2/gallery_of_fluid_motion_final_high.mpg}{high quaility})   
\end{document}